\documentclass[twocolumn,showpacs,floatfix,aps]{revtex4}
\usepackage{graphicx}
\usepackage{dcolumn}
\usepackage{bm}

\begin{document}

\title{Integrability of a disordered Heisenberg spin-1/2 chain}

\author{L. F. Santos}
\email{santos@pa.msu.edu} 
\affiliation{Department of Physics and Astronomy, Michigan State
  University, East Lansing, MI 48824}

\begin{abstract}
We investigate how the transition from integrability to 
nonintegrability occurs by changing
the parameters of the Hamiltonian of a Heisenberg spin-1/2 chain
with defects. Randomly distributed defects may lead to quantum chaos. 
A similar behavior is obtained in the presence of a single 
defect out of the edges of the chain, suggesting that 
randomness is not the cause of chaos in these
systems, but the mere presence of a defect.

\end{abstract}


\maketitle

\section{introduction}

Random matrix theory has long been used to describe the spectra
of complex systems, such as nuclei, molecules, and mesoscopic solids
\cite{weide}. Recently it
has been used in the study of strongly correlated
spin systems \cite{spin,shepe98}. The statistical
properties of the quantum energy spectrum are strongly influenced
by the underlying classical dynamics. The level spacing 
distribution of a classical integrable system is Poissonian,
$P_P(s) = \exp (-s)$, while the level statistics of a chaotic system is 
given by the Wigner-Dyson distribution,
$P_{WD}(s)=(\pi s/2) \exp (-\pi s^2 /4)$. The Wigner-Dyson distribution
is obtained in random matrix theory and it
reproduces the level repulsion of chaotic dynamics.
These two distributions characterize, respectively, the
localized and the metallic phase in the Anderson model of disordered
systems. At the critical point between the two phases an 
intermediate level spacing statistics occurs \cite{inter}.

The problem of localization and the statistical properties
of the spectra for the case of just one particle has long been
understood, but only recently has the 
problem of many-body systems been addressed \cite{spin,shepe98}. 
When more than one particle is present in the system, the 
interaction between them has to be taken into account. The
interplay between interaction and disorder is 
a challenging problem in today's condensed matter physics
and it can lead to new and unexpected effects \cite{us}.

Here, we consider a one-dimensional Heisenberg spin-1/2 chain 
with defects and several excitations.
A defect corresponds to the site where the 
energy splitting is different from all the others. It is 
obtained by applying a different magnetic field in the 
$z$ direction to the chosen site.
In the absence of defects this homogeneous system is integrable and is solved
with the Bethe ansatz \cite{bethe}. Its level distribution is therefore 
Poissonian. As random on-site magnetic fields are turned on and  
their mean-square amplitude starts increasing,
the system undergoes a transition and becomes chaotic. 
But by further increasing the mean-square amplitude, localization
eventually takes place and the distribution becomes Poissonian again.
We determine the crossover from integrability to quantum chaos
in such disordered spin chain. In addition, we discuss
that the cause for nonintegrability is not the 
randomness of the system, but the mere presence of defects. If
only one defect is placed out of the edges of the chain and if the 
defect excess energy is of the order of the 
interaction strength, the system is also chaotic. The same 
sort of transition integrable-chaotic-integrable is obtained as
the defect excess energy increases.

We consider only nearest neighbor interaction.
The Hamiltonian describing the system is

\begin{equation}
H = \sum_{n=1}^{L} (h_n  -\frac{J}{2} \delta _{n,1}
-\frac{J}{2} \delta _{n,L}) \sigma _n^z+
\sum _{n}^{L-1} \frac{J}{4} \vec{\sigma}_{n}.\vec{\sigma}_{n+1},
\end{equation}
where $\hbar =1$ and $\vec{\sigma }$ are Pauli matrices.
There are $L$ sites. 
Each site $n$ is subjected to a magnetic field in the 
$z$ direction, giving the energy splitting $h_n$. 
The chain is ideal whenever all sites have the same energy 
splitting. A defect corresponds to the 
site whose energy splitting differs from the others.

For simplicity, we work with an isotropic chain, 
that is, the coupling constant $J$
for the diagonal Ising interaction
$\sigma_n^z\sigma_{n+1}^z$ is equal to the coupling 
constant for the $XY$-type interaction $\sigma_n^x\sigma_{n+1}^x +
\sigma_n^y\sigma_{n+1}^y$. This last term is responsible for propagating the 
excitation through the chain. A single-particle excitation 
corresponds to a spin pointing up.

In this model, the $z$ component of the total spin $\sum_{n=1}^L S_n^z$ 
is conserved, so 
states with different number of excitations are not coupled. We
therefore look at the level spacing distributions for sectors with the 
same number of excitations. Since we are interested in determining 
if the system is integrable or chaotic, we focus on the sector with 
the largest number of states, that is the sector with $L/2$ excitations. 
This is the region where chaos should set in first.

In a very large system the boundary conditions have no effects,
but numerical calculations are limited to a finite
number of sites. 
In a periodic (or closed) chain we found too many degenerate
states, so we decided to work with a chain with free boundaries 
(or open chain). Both systems, closed or open, 
are known to be integrable in the absence of defects. 
They are solved with the Bethe ansatz method \cite{bethe}.
An open chain with defects 
only on the edges is also integrable \cite{edge}. Here we choose an open
chain with defects of values $-J/2$ on the edges. Such values
nullify the border effects.

\begin{figure}[ht]
\includegraphics[width=3.in]{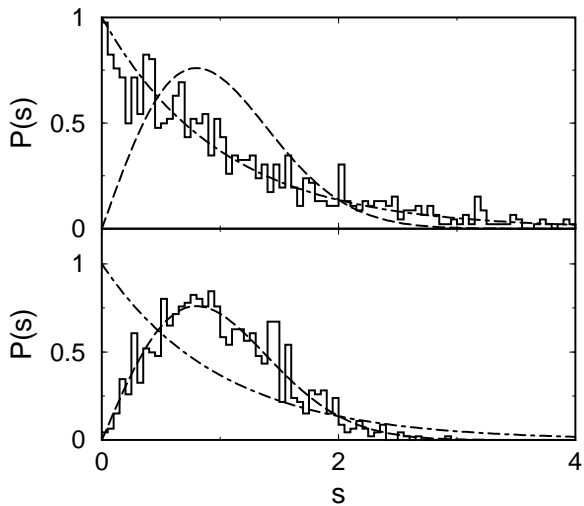}
\caption{In both graphs dot-dashed line
gives the Poisson distribution and long-dashed line corresponds to the 
Wigner-Dyson distribution. The histogram
at the top gives the 
level spacing distribution for a diagonal random matrix of 
dimension 924x924. At the bottom it gives the level distribution 
for a random matrix of this size.}
\label{stat}
\end{figure}

We work with $L=12$ sites and 6 excitations, 
which gives us 924 states. 
A matrix of such size, 924x924, is sufficient to have good statistics, as
illustrated in Fig.~\ref{stat}. In both plots we
have the Poisson distribution (dot-dashed line) and the
Wigner-Dyson distribution (long-dashed line). The spacings $s$ 
correspond to $S/M$, where $M$ is the mean level spacing and $S$ is
the actual spacing. Both histograms are normalized to 1.
The histogram at the top of the figure
shows that the level spacings of the eigenvalues 
of a diagonal random matrix of such size is well described by
the Poisson distribution. The one at the bottom of the figure shows that
the level spacings of the eigenvalues of a random 
matrix of this size agrees very well with the Wigner-Dyson 
distribution. The random elements have a Gaussian distribution.

\section{Randomly distributed defects}

\begin{figure}[ht]
\includegraphics[width=3.in]{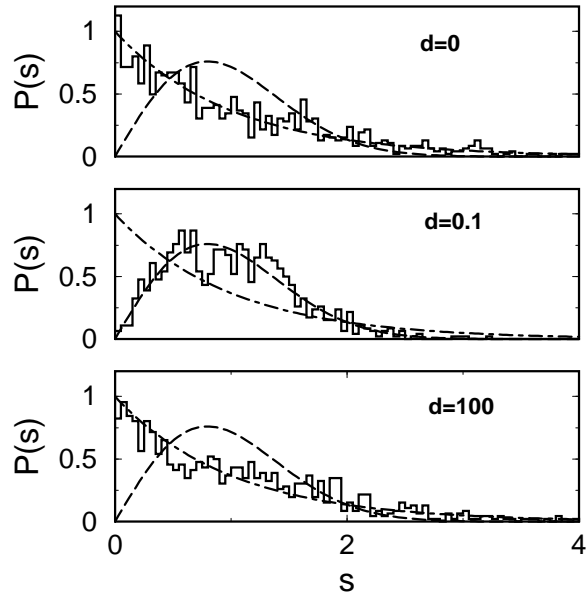}
\caption{The histograms correspond to the actual
level distribution for random on-site magnetic fields. 
We choose $J=1$. Dot-dashed line
gives the Poisson distribution and long-dashed line corresponds to the 
Wigner-Dyson distribution.}
\label{hists_rand}
\end{figure}

First we analyze the case of random magnetic
fields along the $z$ direction. The energy splitting of each site
is given by $h_n = h + d_n$, where 
$d_n$'s 
are uncorrelated random numbers with a Gaussian distribution:
$\langle d_n \rangle =0$ and $\langle d_n d_m \rangle =d^2 \delta _{n,m}$.
When $d=0$ the system is integrable and a Poisson distribution is obtained,
as the histogram at the top of Fig.~\ref{hists_rand} shows. 
As $d$ increases the system undergoes a transition and becomes chaotic, 
the Wigner-Dyson distribution is obtained, as can be seen from the 
histogram in the middle of Fig.~\ref{hists_rand}. However, as we further
increase $d$ and it becomes much larger than $J$,
the system becomes localized. As expected, a 
Poisson distribution reappears (see 
the bottom of Fig.~\ref{hists_rand}). Large $d$ corresponds to a
random diagonal matrix with negligible off diagonal elements.

\begin{figure}[ht]
\includegraphics[width=3.in]{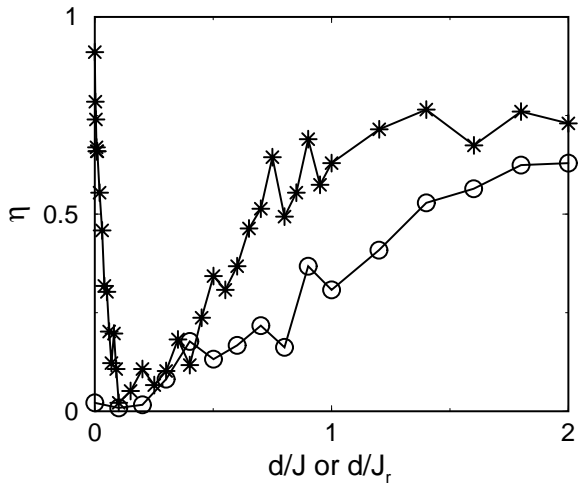}
\caption{Stars give the dependence of $\eta $ on the ratio $d/J$ when 
the on-site magnetic fields are random. Circles give the 
dependence of $\eta $ on the ratio $d/J_r$ when both diagonal and 
non-diagonal elements are random.}
\label{eta_rand}
\end{figure}

A more convenient way to 
analyze the evolution of the level spacing distributions with 
respect to the ratio $d/J$ is by using the parameter $\eta =\int _0^{s_0} 
[P(s) - P_{WD} (s)] ds /\int _0^{s_0} [P_P(s) - P_{WD} (s)] ds$, where
$s_0 = 0.4729...$ is the intersection point of 
$P_P(s)$ and $P_{WD}$ \cite{shepe98,Shepelyansky}. 
A regular system has
$\eta =1 $ and a chaotic system has $\eta =0$. The stars on
Fig.~\ref{eta_rand} show the dependence of $\eta $ on $d/J$. There 
the transition integrable-chaotic-integrable is clear. The 
system is initially integrable, reaches its maximum chaoticity 
when $d\sim 10^{-1} J$ and then becomes integrable again as $d$ becomes
much larger than $J$. At this last step, 
the energy splitting of each site becomes very 
different from all the others and the excitations become localized.

Besides considering the site energies as random numbers, another
way to introduce disorder is by taking the non-diagonal (hopping)
matrix elements at random. We assume that the 
off diagonal elements also have a Gaussian 
distribution and the
mean square is given by $J_r$. Contrary to the case discussed before,
random coupling (circles on Fig.~\ref{eta_rand}) leads to chaos even 
when the energy splittings of all sites are the same ($d=0$). 
Maximum chaoticity is again reached at $d\sim 10^{-1}J_r$, 
but there is now just 
one transition, from chaos to integrability. Such transition 
takes a little longer to happen than in the previous case, though
localization is also attained once 
the ratio $d/J_r$ becomes large. 

\section{One single defect}

\begin{figure}[ht]
\includegraphics[width=3.in]{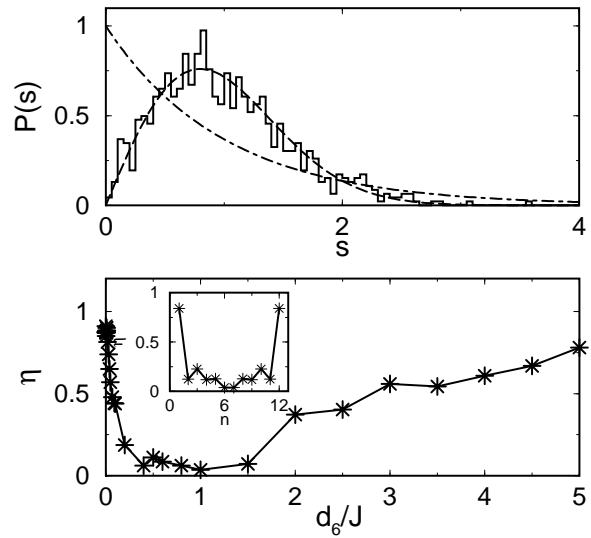}
\caption{Top panel: The histogram corresponds to the level distribution 
of the Heisenberg spin-1/2 chain with a defect on site 6. The 
defect excess energy is 
equal to the interaction strength $J$. Dot-dashed and 
long-dashed lines give 
the Poisson and the Wigner-Dyson distributions respectively.
Bottom panel: Dependence of the parameter $\eta $ on the 
ratio $d_6/J$. The defect is placed on site 6. The inset
gives the dependence of $\eta$ 
on the position of the defect; the defect excess energy is $J$.}
\label{def6_all}
\end{figure}

In the system treated here, chaos can be associated with randomness 
only when the coupling is random. 
In the case of a constant interaction strength,
what is really responsible for the nonintegrability of the 
system is not the randomness of the energy splittings, 
but the mere existence of 
a defect. Let us consider again a constant coupling.
The top of Fig.~\ref{def6_all} 
shows that a Wigner-Dyson distribution can also be obtained
when there is only one defect in the middle of the chain
and the defect excess energy is of the order of the interaction 
strength. For this histogram all sites have the same energy splitting 
$h_n = h$, except site 6, which has $h_6=h+J$.

Here too we use the
parameter $\eta $ to study the evolution of the 
level spacing distribution with respect to the ratio $d_6/J$. 
The bottom of Fig.~\ref{def6_all}
indicates a transitional behavior very similar to the case of
constant coupling and random diagonal elements.
The one-defect system is initially regular, but by increasing $d_6$ 
it becomes chaotic.
The highest degree of chaoticity 
happens when $d_6 \sim J$. The level 
repulsion now settles in more slowly than in the previous situation
of random magnetic fields. As $d_6$ 
is further increased, an excitation on site 6 will
become site-localized. This means that when the ratio 
$d_6/J$ becomes very large there are two kinds of states in the system:
states with a localized excitation on site 6 and states with no excitation
on the defect. These two types of states are not coupled. This system
becomes equivalent to two smaller and uncoupled ideal chains, 
and integrability is therefore recovered.

In the inset at the bottom of Fig.~\ref{def6_all},
we again consider one single defect,
whose excess energy is 
equal to $J$. It shows how the parameter $\eta $ depends on the position
of the defect and
confirms that 
the Heisenberg spin-1/2 system is integrable when it has 
defects placed on its edges.

\section{Conclusion}

We have shown that in a Heisenberg spin-1/2 chain, randomly distributed
defects lead to the transition integrable-chaotic-integrable, according
to the ratio $d/J$. When random off diagonal elements are also considered, 
there is only one transition from non-integrability to integrability. 
The transition integrable-chaotic-integrable is also observed when 
the coupling is constant and there is only 
one defect out of the edges of the chain. 
The level spacing distribution obtained in this case depends on how
large the defect excess energy is in relation to the interaction strength.

Obtaining analytical solutions for disordered spin chains
is not an easy task and in many cases is simply impossible.
The algebraic version of the Bethe ansatz is often used to 
{\it construct} integrable Hamiltonians \cite{Zvy}.
The analysis of level spacing distributions should
therefore be useful to identify which real or constructed Hamiltonians 
are indeed integrable.

Understanding under what conditions
disordered spin systems become integrable is not just
relevant for condensed matter physics, but also for quantum computation,
since these systems 
are commonly used to model different proposals of quantum computers 
(see Ref.~\cite{us} and references therein).

\acknowledgments 
We acknowledge support by the NSF through grant No. ITR-0085922 
and would also like to thank C. O. Escobar for helpful discussions.

\end{document}